\documentclass[aps,prl,twocolumn,nofootinbib]{revtex4-2}

\usepackage{preamble}

\begin{document}

\title{Gravitational-wave signatures of nonviolent nonlocality}

\author{Brian C. Seymour\,\orcidlink{0000-0002-7865-1052}} %
\email{seymour.brianc@gmail.com}
\author{Yanbei Chen\,\orcidlink{0000-0002-9730-9463}} %
\affiliation{TAPIR, Walter Burke Institute for Theoretical Physics, California Institute of Technology, Pasadena, California 91125, USA}

\date{\today}

\begin{abstract} 
    Measurement of gravitational waves can provide precision tests of the nature of black holes and compact objects. In this work, we test Giddings' nonviolent nonlocality proposal, which posits that quantum information is transferred via a nonlocal interaction that generates metric perturbations around black holes. In contrast to firewalls, these quantum fluctuations would be spread out over a larger distance range --- up to a Schwarzschild radius away. In this letter, we model the modification to the gravitational waveform from nonviolent nonlocality. We modify the nonspinning EOBNRv2 effective one body waveform to include metric perturbations that are due to a random Gaussian process. We find that the waveform exhibits random deviations which are particularly important in the late inspiral-plunge phase. We find an optimal dephasing parameter for detecting this effect with a principal component analysis. This is particularly intriguing because it predicts random phase deviations across different gravitational wave events, providing theoretical support for hierarchical tests of general relativity. We estimate the constraint on the perturbations in nonviolent nonlocality with events for the LIGO-Virgo network and for a third-generation network.
\end{abstract}
\maketitle

\noindent \textit{\textbf{Introduction---}} The evaporation of black holes (BHs) via Hawking radiation \cite{Hawking:1975vcx} reveals an inconsistency between quantum mechanics and general relativity (GR) \cite{Hawking:1976ra}: the semiclassical result that Hawking radiation carries no information contradicts the {\it unitarity} of quantum theory. Several resolutions to the {\it information paradox} have been proposed: the BH could never fully decay but remain as a massive remnant \cite{Giddings:1992hh}. The interior geometry of the BH could be modified as a fuzzball \cite{Mathur:2005zp} or gravastar \cite{Mazur:2004fk}. Alternatively, the firewall scenario \cite{Almheiri:2012rt} suggests the region near the horizon could experience a breakdown of semiclassical gravity which destroys infalling observers. Finally, the information paradox may be resolved by accounting for nonperturbative contributions in semiclassical gravity through the replica wormhole trick, which restores unitarity without modifying GR \cite{Almheiri:2019psf,Almheiri:2019hni,Penington:2019kki,Almheiri:2019qdq, Almheiri:2020cfm, Geng:2024xpj}.

Giddings proposed that the information which fell into the BH can escape it via {\it nonviolent nonlocality} (NVNL)~\cite{Giddings:2011ks,Giddings:2012gc,Giddings:2014nla,Giddings:2017mym,Giddings:2024qcf}, a nonlocal interaction between the inside and the outside of the BH, with associated nonviolent space-time fluctuations from this information transfer. In the {\it strong} version of NVNL, the space-time metric fluctuates stochastically at a level of $\mathcal{O}\left( 1 \right)$ \cite{Giddings:2014nla} near BHs, while the {\it weak} version \cite{Giddings:2017mym} has fluctuations of $\mathcal{O}\left( e^{-S_\mathrm{bh}/2} \right) $ \cite{Giddings:2022jda}. Giddings and collaborators elaborated the phenomenology of NVNL~\cite{Giddings:2014ova,Giddings:2016tla,Giddings:2016btb,Fransen:2024fzs} to observations of the Event Horizon Telescope (EHT) \cite{EventHorizonTelescope:2022wkp}. 

Gravitational wave (GW) measurements directly probe the behavior of strong gravity around BHs and neutron stars (NSs)~\cite{LIGOScientific:2016aoc, LIGOScientific:2017ycc, LIGOScientific:2018dkp,LIGOScientific:2019fpa,LIGOScientific:2020tif,LIGOScientific:2021sio}. Deviations from GR can be extracted by adopting either the parameterized post-Einsteinian (PPE) framework~\cite{Yunes:2009ke} or the post-Newtonian (PN) deformation framework~\cite{Li:2011cg,Agathos:2013upa, Mehta:2022pcn}, which measure deviations in the GW phase during the binary inspiral stage. Results of parametrized tests have been featured in LIGO-Virgo-KAGRA (LVK) results~\cite{LIGOScientific:2016lio,LIGOScientific:2018dkp,LIGOScientific:2019fpa,LIGOScientific:2020tif,LIGOScientific:2021sio}, which also treated the phase deviation parameters as hyperparameters and bounded their mean values as well as uncertainties~\cite{Isi:2019asy, Saini:2023rto}. In this Letter, we model the effect of NVNL on the inspiral of a binary black hole (BBH) and show that it can be constrained by the parametrized tests mentioned above, in particular via the uncertainties of the phase-deviation hyperparameters~\cite{Isi:2019asy}.

We begin with the effective-one-body (EOB) framework \cite{Buonanno:1998gg,Buonanno:2000ef,Damour:2001tu} which approximates a two-body relativistic problem as a one-body problem in a deformed Schwarzschild spacetime, the {\it effective} spacetime. Originally, this was developed for nonspinning quasicircular \cite{Buonanno:1998gg,Buonanno:2000ef,Damour:2001tu,Pan:2011gk,Barausse:2011dq,Damour:2016bks} but subsequently was generalized to aligned spin \cite{Pan:2009wj,Taracchini:2012ig,Bohe:2016gbl,Nagar:2020pcj}, precessing spin \cite{Damour:2008qf,Barausse:2009xi,Barausse:2011ys,Pan:2013rra,Taracchini:2013rva,Nagar:2018zoe}, calibrated to numerical relativity \cite{Buonanno:2007pf,Buonanno:2009qa,Babak:2016tgq}, tides \cite{Damour:2009wj, Baiotti:2011am,Steinhoff:2016rfi,Hinderer:2016eia}, and eccentricity \cite{Cao:2017ndf,Ramos-Buades:2021adz} where the newest model is SEOBNRv5PHM \cite{Pompili:2023tna,Khalil:2023kep,Ramos-Buades:2023ehm,Haberland:2025luz}. We modify a simple nonspinning EOBNRv2 spacetime~\cite{Pan:2011gk} and add stochastic metric perturbations characterized by a Gaussian spatial profile and a frequency spectrum related to the BH's temperature, as proposed by Ref.~\cite{Giddings:2016btb}. The resulting trajectories lead to GWs that deviate from GR stochastically, primarily in the late-inspiral and plunge phases. Using a principal component analysis (PCA) \cite{Pai:2012mv,Saleem:2021nsb,Datta:2022izc}, the frequency-domain phase deviations of NVNL waveforms are well approximated by a single dominant eigenmode multiplied by a normally distributed random amplitude. Finally, we estimate how well NVNL can be constrained by stacking together gravitational-wave events and computing the Savage-Dickey ratio \cite{Dickey:1971}. These calculations give theoretical support to search for random phase deviations as proposed in the hierarchical tests of GR \cite{Isi:2019asy,LIGOScientific:2021sio,Isi:2022cii,Payne:2023kwj,Payne:2024yhk,Zhong:2024pwb}.

\noindent \textit{\textbf{Setup---}} For a Schwarzschild BH, Ref.~\cite{Giddings:2016btb} generically decomposes the NVNL-induced metric perturbations $g_{\mu\nu} = g_{\mu\nu}^\mathrm{s} + n_{\mu\nu}$ into the even and odd perturbations~\cite{Regge:1957td}. In this work, we will consider the dominant NVNL correction which arises from $n_{vv}$ in the in-going Eddington-Finkelstein coordinates, decomposed as
\begin{equation}
    n_{vv} = \sum_{\ell m}f_{\ell m} Y_{\ell m} \, ,
\end{equation}
where 
\begin{equation}
    f_{\ell m} = A_{\ell m} \exp\left[ -{\left( r-r_S \right)^2}/{(2 r_G^2)} \right] n(t)\, ,
\end{equation}
with $A_{\ell m}$ the amplitude of the mode, $r_S$ the  Schwarzschild radius, and $r_G\sim r_S$ the localization length of the perturbations \cite{Giddings:2016btb}. Here $n(t)$ is a colored Gaussian noise with a power spectrum 
\begin{equation}
\label{eqSn}
    S_n(f)=1/({2 f_Q})\exp\left[ -\vert f\vert/f_Q \right]\, ,
\end{equation}
where $f_Q = 1/8\pi M$ is the quantum frequency scale. We have normalized the power spectrum so that $n(t)$ has a variance equal to unity $\langle n^2(t)\rangle \equiv \int df S_n(f) =1$. This spectrum is motivated by the Boltzmann distribution for a BH of temperature $T_\mathrm{BH} = 1/8\pi M$. Note that the standard deviation of $f_{\ell m}$ at the horizon is $\sim A_{\ell m}$, while the coherence time is around $\tau \sim 4 M$.

For a test particle falling into a NVNL BH, using $\epsilon$ and $\eta$ as expansion parameters in particle mass and size of NVNL, we can perform a two-parameter expansion for the metric as $g = g^\mathrm{S}+ \eta n +\epsilon h^{(0,1)} + \eta \epsilon h^{(1,1)}$, and the trajectory of the particle as $x^{(0,0)}+\eta x^{(1,0)}$. The modified Einstein equation, up to $O(1,1)$ can be written as $G[g+\eta n] = 8\pi \eta T_{\rm NVNL}$ and 
{\small 
\begin{equation}\label{modified:einstein}
G^{(1)}_{g+\eta n}[\epsilon h^{(0,1)}+\epsilon \eta h^{(1,1)}] =8\pi \epsilon T_{g+\eta n}[x^{(0,0)}+\eta x^{(1,0)}] 
\end{equation}
}where $G^{(1)}_{g+\eta n}$ denotes the linearized Einstein tensor with background $g+\eta n$ and $T_{g+\eta n}[x]$ denotes stress-energy tensor of a particle in this background with world line $x$. From Eq.~\eqref{modified:einstein}, we recover $G^{(1)}_g [h^{(0,1)}] =8\pi T_g[x^{(0,0)}]$, obtain that $x^{(0,0)}+\eta x^{(1,0)}$ is a geodesic in $g +\eta n$ (linear Bianchi identity), and that 
{\small 
\begin{equation}\label{eqh11}
    G_g^{(1)}[h^{(0,1)}+ \eta h^{(1,1)}] = 8\pi T_{g+\eta n}[x^{(0,0)}+\eta x^{(1,0)}] -\eta G_g^{(2)}[h^{(0,1)},n] \,.
\end{equation}
}Equation~\eqref{modified:einstein} indicates that GWs up to $O(1,1)$ are generated by a modified trajectory and propagate on the modified spacetime $g+\eta n$; Eq.~\eqref{eqh11} shows further that the propagation effect in $h^{(1,1)}$ can be computed as being sourced by a beat between $h^{(0,1)}$ and the metric perturbation $n$. Within GR, radiation reaction effects show up in $x^{(0,1)}$ and $h^{(0,2)}$, while NVNL corrects in $x^{(1,1)}$ and $h^{(1,2)}$. In this Letter, we shall restrict ourselves to the test-particle limit since, as we see in Supplemental Material \cite{supp}, the NVNL is confined near the end of the inspiral and the finite-mass effects will not cumulate secularly. We will further neglect the wave propagation effect, since that involves the scattering of the merger wave and causes a modified ringdown wave --- while we will be dealing with inspiral tests for NVNL.

Let us now review the EOB framework \cite{Buonanno:1998gg,Buonanno:2000ef,Damour:2007cb,Damour:2009kr,Buonanno:2009qa,Pan:2011gk}. We define total mass $M = m_1 + m_2$, symmetric mass ratio $\eta = m_1 m_2 / M^2$, and mass ratio $q = m_1/m_2$. The effective metric is given by
\begin{equation}
    ds^2_\mathrm{eff} = - A(r) dt^2 + \frac{D(r)}{A(r)} dr^2 + r^2 d\Omega^2 \, ,
\end{equation}
where $A(r) = 1 - \frac{2 M}{r} + \mathcal{O}(\eta)$ and $D(r)=1+\mathcal{O}(\eta)$ are terms that describe the effective metric which is deformed by the symmetric mass ratio $\eta$ away from Schwarzschild \cite{Buonanno:1998gg,Buonanno:2000ef}. To capture the order of magnitude of the NVNL effect, let us now add NVNL into EOB by modifying the geometry of the effective spacetime, the modified mass-shell relation $p_\mu p_\nu \left( g^{\mu\nu}_\mathrm{S} + n^{\mu\nu} \right) = - 1$ then yields a modified Hamiltonian
\begin{equation}
    \hat H_\mathrm{real} = \hat H_\mathrm{real}^\mathrm{S} + n_{vv}^{\ell m} \, \Delta \hat H_{\ell m}^\mathrm{real} \, .
\label{eq:HrealNVNL}
\end{equation}
In principle, the GW luminosity $\dot E$ is also modified by NVNL~\cite{Fransen:2024fzs} (and correspondingly $\mathcal{F}^{\rm rad}_i$), but here we focus on the conservative modifications. We obtain the NVNL trajectory using Eq.~\eqref{eq:HrealNVNL}, then the leading quadrupole wave $h_{22}$ in the same way as Ref.~\cite{Pan:2011gk}, attaching a simple GR ringdown where the strain peaks via smoothness (similar to Ref.~\cite{Pompili:2023tna}). (See Supplemental Material for details which includes Refs.~\cite{Husa:2015iqa,Khan:2015jqa}.)

\begin{figure}[t]
    \centering
    \includegraphics{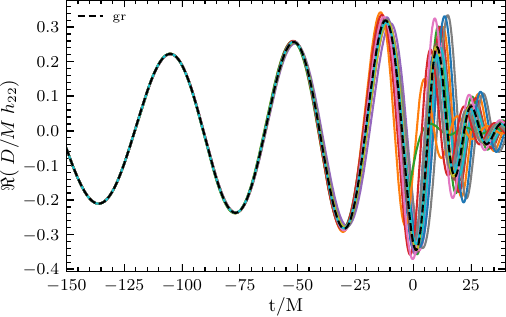}
    \caption{The real part of the dimensionless $h_{22}(t)$ strain for $A_{22} = 5\times 10^{-2}$. This is the \textit{full waveform} before the PCA is done, so it contains all perturbations. These signals are aligned at very early times so that their signals overlap at low frequencies but they stochastically diverge as they reach the plunge. The apparent ringdown difference is primarily due to the phenomenological ringdown attachment, but we only do the testing GR analyses with the inspiral piece.}
    \label{fig:td-wf}
\end{figure}

\noindent \textbf{\textit{NVNL Waveforms---}}
In Fig.~\ref{fig:td-wf}, we plot several realizations of an NVNL waveform $\mathrm{Re}[h_{22}]$ (normalized by $M/D$ with $D$ the source distance) in the time domain, as functions of $t/M$, for a binary with $q=1$, and compare this to a GR waveform. The random deviations are smooth in time due to the cutoff $f_Q$ in Eq.~\eqref{eqSn} as well as the filtering effect due to the inertia of
the binary which suppresses high frequencies. 

\begin{figure}[th]
    \centering
    \includegraphics{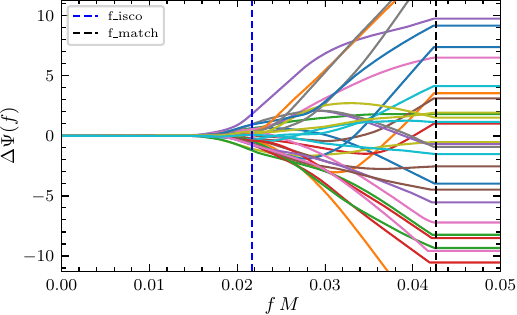}
    \caption{Frequency domain phase deviation realizations for $A_{22} = 1$. Using time domain waveform realizations shown in Fig.~\ref{fig:td-wf}, we plot the amount of dephasing from GR that they will have. We also plot the frequency at which the binary crosses the inner most stable circular orbit (dashed blue) and the frequency at which the inspiral portion of the waveform is matched to the ringdown (dashed black). Note that $A_{22} = 1$ is not a small deviation from GR, so we calculated this at $A_{22} \ll 1$ and scaled it appropriately. One can see that the secular effect of NVNL is nearly zero while the theory predicts random dephasing from GR.}
    \label{fig:phase-shift}
\end{figure}

Going to the frequency domain, the NVNL waveform at linear order in $A_{\ell m}$ can be approximated by
\begin{equation}\label{eq:general-wf-deviation}
    h(f; \theta, A_{\ell m}) = h_\mathrm{GR}(\theta) e^{i \sum_{\ell m} \Delta  \Psi_{\ell m}(f;\theta) } \, ,
\end{equation}
with $\Delta \Psi_{\ell m}(f)\propto A_{\ell m} $ a stochastic phase deviation from GR, found by simulating a NVNL waveform with metric fluctuations $n_i(t)$ in the time domain and taking the Fourier transform for a particular noise realization. 

Since our metric deviations are Gaussian, the deviation in the frequency domain is just the metric deviations multiplied by a transfer function. In Fig.~\ref{fig:phase-shift}, one can see the frequency domain phase deviations for various noise realizations. Notice that the frequency domain phase is primarily a stochastic deviation rather than having a secular effect that is common to all these noise realizations. Let us define the quantity $\mu_{\ell m} \equiv \langle\Delta \Psi_{\ell m}(f)\rangle$ and $\Sigma_{\ell m}(f,f') = \langle \left[ \Delta \Psi_{\ell m}(f) - \mu_{\ell m}(f) \right] \left[ \Delta \Psi_{\ell m}(f') - \mu_{\ell m}(f') \right] \rangle $. The mean deviation turns out to be negligible, and detection hinges on finding the presence of $\Sigma_{\ell m}$.

\begin{figure}[t]
    \centering
    \includegraphics{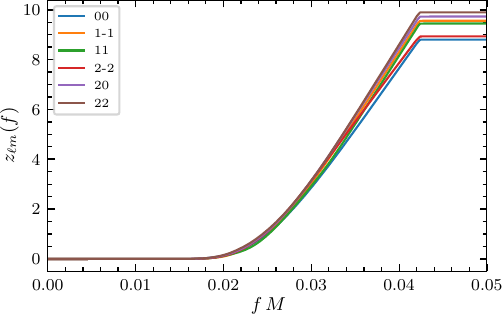}
    \caption{The largest PCA modes of the covariance matrix. All modes with odd $l+m$ are zero since we are confined to the orbital plane with $\theta = \pi/2$. The largest eigenvector accounts for $\sim 97\%$ of the phase variance for each of the modes. One can see that similar deviations happen for all $(\ell, m)$ modes.}
    \label{fig:eigenvectors-lanczos-modes}
\end{figure}

We can simplify the waveform model Eq.~\eqref{eq:general-wf-deviation} by performing a PCA, which corresponds to diagonalizing the covariance matrix
\begin{equation}\label{eq:PCA}
    \Sigma_{\ell m}(f,f') = \sum_k \left( \sigma_{\ell m}^k \right)^2 z^k_{\ell m}(f) z^k_{\ell m}(f') \, ,
\end{equation}
where $(\sigma_{\ell m}^k)^2$ is the $k$th largest eigenvalue and $z^k_{\ell m}(f)$ its corresponding eigenvector. The PCA is performed for $f_\mathrm{start}< f < f_\mathrm{match}$ where $f_\mathrm{start} = 0.004$ and $f_\mathrm{match} = 0.042/M$\footnote{Note that the output of the dominant PCA eigenvector is completely equivalent to finding $\min \limits_z \sum_{ij} \left(\Sigma(f_i,f_j) - z(f_i) z(f_j)\right)^2$ with constant norm.}. As it turns out, for a nearly equal mass ratio, using a single component $k=0$ is able to capture more than 97\% of the total variance for the $(2,2)$ mode, similarly for other $(\ell,m)$ modes. We shall then adopt
\begin{equation}\label{eq:pca-equation-phase}
    h(f; \theta, A_{\ell m}) = h_\mathrm{GR}(\theta) e^{i \sum_{\ell m} \zeta_{\ell m} z_{\ell m}(f) } \, ,
\end{equation}
where $\zeta_{\ell m} \sim \mathcal{N}(\mu_{\ell m}, \sigma_{\ell m})$ and we dropped the $k=0$ index; we further scale the eigenvectors so that $\sigma_{\ell m}=A_{\ell m}$, and note that $\mu_{\ell m}/\sigma_{\ell m} \ll 1$.
In Fig.~\ref{fig:eigenvectors-lanczos-modes}, we plot $z_{\ell m}(f)$ for each mode up to $l = 2$. Since these curves are close to each other, we will only include $(\ell,m) = (2,2)$ and neglect the $(\ell,m)$ labeling henceforth. We stress that we apply PCA directly to the NVNL theory itself, rather than in the measurement space, which is the usual approach in the literature \cite{Pai:2012mv, Saleem:2021nsb, Datta:2022izc}. See Supplemental Material for further details of the PCA.

\noindent \textit{\textbf{Extraction of NVNL from data---}} Next, we describe how well hyperparameters $(\mu,\sigma)$ can be estimated from a collection of GW events with $\zeta$ drawn from a true distribution $\mathcal{N}(\mu_\true=0,\sigma_\true=A)$. For events with high signal-to-noise ratio (SNR), parameter estimation accuracy can be quantified by the Fisher information matrix \cite{Finn:1992wt}:
\begin{equation}
    \Gamma_{IJ} = \left.\left( \partial_I h\mid \partial_J h \right)\right|_{\Theta=\Theta_\true} \, .
\end{equation}
We collect all our signal parameters into an uppercase-indexed vector $\Theta^I = \left( \theta^i,\zeta \right)$ where lower case index $i = 1,\dots, n_\mathrm{\Theta}-1$ run over the standard GR parameters $\theta^i$ and the extra entry $\zeta$ is the NVNL parameter. The capital letters range over $I = 1, \dots, n_\mathrm{\Theta}$. The noise-weighted inner product is defined as 
\begin{equation}
    (g\mid h)=4 \operatorname{Re} \int_0^{\infty} \frac{\tilde{g}^*(f) \tilde{h}(f)}{S(f)} d f \, ,
\end{equation}
Given data $d$ (which contains a high SNR signal) from a single event, under the Fisher approximation, the likelihood function is taken to be
\begin{equation}
    p(d|\Theta)=  \sqrt{\frac{\det \Gamma}{\left( 2\pi \right)^{n_\Theta} }} \exp \left[-\frac{1}{2}( \Theta_I - \Theta_I^\ml )\Gamma_{IJ}( \Theta_J - \Theta_J^\ml ) \right] \, ,
\end{equation} 
where we have used $\Theta_J^\ml$ to denote the values of $\Theta_J$ where likelihood is maximized and $n_\Theta$ is the number of entries of $\Theta_I$. In a frequentist approach, we can use $\Theta_I^\ml$ as the maximum likelihood estimator (MLE) for signal parameters. Given a large number of trials with true parameters $\Theta_I^\true$, we can denote $\delta\Theta_I = \Theta_I^\ml-\Theta^\true_I$.  For high SNR, $\delta\Theta_I$ is a Gaussian random vector with $\langle \delta\Theta_I\delta\Theta_J\rangle = (\Gamma^{-1})_{IJ}$.  In particular, marginalizing over $\theta_i$, the MLE estimator $\zeta^\ml$ has an error of $\langle\delta\zeta^2\rangle = (\Gamma^{-1})_{\zeta\zeta} \equiv \Delta\zeta^2$.

Even though NVNL is best represented by its principal component in Eq.~\eqref{eq:pca-equation-phase}, LVK's PN deformation tests~\cite{Yunes:2009ke, Li:2011cg, Agathos:2013upa} can still search for NVNL with reasonable efficiency. Standard techniques for biased waveform models~\cite{Cutler:2007mi,Hu:2022bji} will be applied in Ref.~\cite{Seymour:2026bjg}, with an abbreviated derivation provided here. We adopt a phase deviation of Ref.~\cite{LIGOScientific:2021sio} (with a single $n$):
\begin{equation}
    \Delta\Psi_n(f) =  \frac{3}{128\eta}  \varphi_n \delta \varphi_n \left( \pi M f\right)^{(n-5)/3} \, .
\end{equation}
If we inject a $\zeta^\true$, the MLE for $\Theta^I = (\theta^i,\delta\varphi_n)$ is given by
{\small 
\begin{equation}\label{eq:biasedpn}
    \Gamma_{IJ} 
    \begin{pmatrix}
        \theta_i^\ml - \theta_i^\true \\
        \delta\varphi_n^\ml
    \end{pmatrix} = \begin{pmatrix}
        \left(\partial_{\theta^i} h| i \Delta\Psi_\mathrm{NVNL} h_\mathrm{GR}\right) \\
        \left(\partial_{\delta\varphi_n} h| i \Delta\Psi_\mathrm{NVNL} h_\mathrm{GR}\right)
    \end{pmatrix}+\begin{pmatrix}
        \left(\partial_{\theta^i} h|n\right) \\
        \left(\partial_{\delta\varphi_n} h| n\right)
    \end{pmatrix} \, ,
\end{equation}
}where $h$ is evaluated at $(\theta_\true^i,\delta\varphi=0)$ and the maximum likelihood values of $\theta^\ml_i$ and $\delta\varphi_n^\ml$ are implicitly defined in Eq.~\eqref{eq:biasedpn}. We are expanding about small $\delta\varphi_n^\ml $ and $\theta^\ml_i - \theta_\true^i$ and using Eq.~(10) of Ref.~\cite{Cutler:2007mi}.
This means that $\delta \varphi_n^\ml = \Sigma_{\delta\varphi_n, J} \left(\partial_J h |i z(f) h_\mathrm{GR}\right)\zeta^\true$ with an uncertainty with covariance matrix $\Sigma_{IJ} = \left( \Gamma^{-1} \right)_{IJ}$. Thus for a small bias, the maximum likelihood point is shifted, but the statistical uncertainty is given by the Fisher matrix calculated with the parameterized test parameters $(\theta_i,\delta \varphi_n)$. 

Let us now construct a hierarchical analysis for the distribution of $\zeta$ for a collection of events, similar to Ref.~\cite{LIGOScientific:2021sio}. We model $\zeta$ as $\zeta\sim \mathcal{N}(\mu,\sigma)$ and would like to estimate the posterior on the hyperparameters $(\mu,\sigma)$. For event $a$, we write
\begin{align}
    p(d_a|\mu,\sigma) &=  \int d\zeta p(d_a|\zeta) p(\zeta| \mu,\sigma) \, , \nonumber\\
    &=  \frac{1}{\sqrt{2\pi} \sqrt{\Delta\zeta_a^2 + \sigma^2}} \exp \Big[ - \frac{1}{2} \frac{ (\zeta_a^\ml -\mu)^2}{\Delta\zeta_a^2 + \sigma^2} \Big] \, ,
\end{align}
where $\zeta_a^\ml$ and $\Delta\zeta_a$ are the MLE and parameter uncertainty for $\zeta$ obtained from this event (and thus depend on $d_a$). Note that the maximum likelihood point for event $a$ has the distribution
\begin{equation}
    \zeta_a^\ml \sim \mathcal{N}\left( 0, \sqrt{\sigma_\true^2 + \Delta\zeta_a^2} \right) \, ,
\end{equation}
which follows from $\zeta^\true_a \sim \mathcal{N}\left( 0, \sigma_\true\right)$ and $\zeta_a^\ml \sim \mathcal{N}\left( \zeta_a^\true,\Delta\zeta_a \right)$. If a generic PN test were performed instead, the maximum likelihood point is distributed like
\begin{equation}
    \delta \varphi_{n,a}^{\ml} \sim \mathcal{N}\left(0,\sqrt{ \alpha_a^2 \sigma_\true^2 + \left( \Delta \delta\varphi_{n,a} \right)^2}  \right) \, ,
\end{equation}
since $\delta\varphi^\ml_{n,a} \sim \mathcal{N}\left( \alpha_a \zeta^\true,  \Delta \delta\varphi_{n,a}\right)$ where $\alpha_a \equiv \Sigma_{\delta\varphi_n,I} \left(\partial_I h | i z(f) h_\mathrm{GR} \right)$ is the coupling for event $a$ given by Eq.~\eqref{eq:biasedpn} and $\Delta \delta \varphi_{n,a}$ is the statistical uncertainty on it. The entire collection of events leads to the joint likelihood
\begin{equation}
    p(\left\{ d_a \right\}|\mu,\sigma) = \prod_{a=1}^N p(d_a|\mu,\sigma) \, ,
\end{equation}
where $(\mu,\sigma)$ are the hyperparameters for $\zeta$, but the framework works analogously for PN deformation hyperparameters $(\mu_n,\sigma_n)$. The number of events is $N$, and we do not include corrections for selection effects \cite{Magee:2023muf} or the probability of obtaining $N$ events \cite{Thrane:2018qnx,Payne:2023kwj}. To compute the consistency with GR, we use Bayes factors which compare the support for or against GR. Since we are comparing a nested model where GR is a single point $(0,0)$ in the $(\mu,\sigma)$ plane, it becomes the well-known Savage-Dickey ratio \cite{Dickey:1971}. The Bayes factor $\mathcal{B}$ is defined to be the ratio of the evidences 
\begin{equation}
    \log \mathcal{B}^\mathrm{bGR}_\mathrm{GR} = \log \left( \frac{p(d|\mathrm{bGR})}{p(d|\mathrm{GR})} \right) = \log \left( \frac{p(0,0|d,\mathrm{bGR})}{p(0,0|\mathrm{bGR})} \right) \, ,
\end{equation}
where we are using the notation $p(x|d,M)$ to represent the posterior probability density $x$ given data $d$ under modeling assumptions $M$ and ``bGR'' denotes for the beyond GR model. One can see that the Savage-Dickey ratio compares how much the posterior has changed to the prior at the location of the GR limit. We use priors which are uniform in the range of $-1\leq \mu\leq 1$ and $0\leq\sigma\leq 1$.

\begin{figure}
    \centering
    \includegraphics{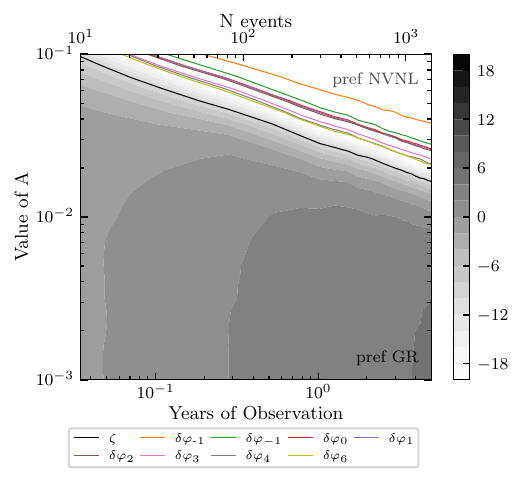}
    \caption{The log Bayes factor projected constraint for Hanford-Livingston-Virgo network operating at O3 Livingston sensitivity (positive favors GR). We plot this for various values of $A$ and for increasing numbers of events. The line corresponding to $-10$ log Bayes factor is shown for the optimal PCA model (black) and PN coefficients (other colors) where GR is disfavored. Using an event list, we perform parameter estimation for five years of detectable events and then compute the Bayes factor for the hierarchical test of GR. Note that the PCA model is best able to constrain the effects of NVNL most stringently, but the PN coefficients are able to detect a violation of $A\ne0$ nearly as well. We also see that the largest PN orders perform the best. For a five year observation, the bound for the PCA model is $A < 1.6\times 10^{-2}$.}
    \label{fig:bayes-factor-result-O3}
\end{figure}

\begin{figure}
    \centering
    \includegraphics{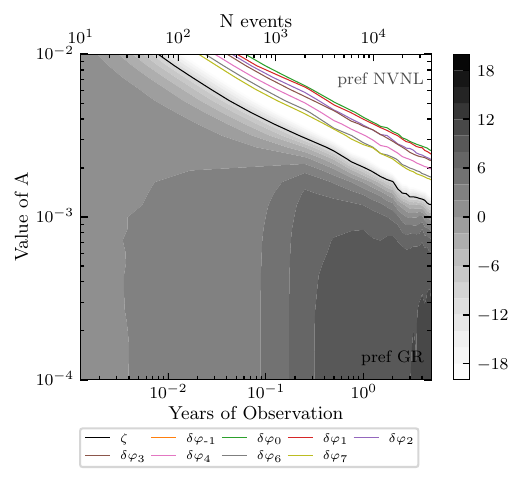}
    \caption{
    The log Bayes factor for the Hanford-Livingston-Virgo network operating at CE sensitivity (positive favors GR). This plot shows the same scaling as Fig.~\ref{fig:bayes-factor-result-O3}, but contains more events since CE detects more in a five year period. For a five year observation, the bound for the PCA model is $A < 1.2\times 10^{-3}$.
    }
    \label{fig:bayes-factor-result-CE}
\end{figure}

Next, we investigate the detectability of NVNL via signal injection. In Fig.~\ref{fig:bayes-factor-result-O3} we show a contour plot of the Savage-Dickey ratio for a three detector network Livingston-Hanford-Virgo at O3 Livingston sensitivity. In particular, we calculate the Fisher information matrix for five years of events where we draw from the astrophysical rates. The merger rate density scales with the star formation rate and the masses are drawn from the \textsc{Power Law+Peak} (PP) model \cite{Fishbach:2018edt,Talbot:2018cva} that is the best fit point from GWTC3 data \cite{KAGRA:2021duu}. We show the estimated measurement precision in the astrophysical parameters, marginalizing over the event-level parameters. We perform this analysis for the optimal PCA dephasing term. Additionally, we give the results for the traditional PN deformation coefficients based upon the biased framework \cite{Seymour:2026bjg}.
We can see that NVNL can be constrained such that $A \lesssim 1.6\times 10^{-2}$ after five years of observation at O3 sensitivities when using the PCA method $\zeta$. The 3.5PN dephasing term, $\delta\varphi_7$, performed the best out of the parameterized tests and found the constraint $A \lesssim 2.1\times 10^{-2}$. We note that for a single BBH event $A\lesssim0.1$ which is comparable with what Ref.~\cite{Liebling:2017pqs} found using a different scheme (cf.~Table 1 \cite{Liebling:2017pqs}). The minimum detectable value depends weakly on PN order, but the PCA method was optimal. This is consistent with previous work showing that deviations are detectible with most PN tests \cite{Sampson:2013lpa,Meidam:2017dgf}. For third-generation detectors, we consider three Cosmic Explorer detectors located at the locations of the current LIGO-Virgo network. The five-year constraint on $A$ is now $A \lesssim 1.2\times 10^{-3}$ for the PCA method and $A \lesssim 1.7\times 10^{-3}$ for the 3.5PN parameterized test, as shown in Fig.~\ref{fig:bayes-factor-result-CE}. For Figs.~\ref{fig:bayes-factor-result-O3} and \ref{fig:bayes-factor-result-CE}, we resampled many realizations of events and their associated noise realization so that very loud events do not show drastic shifts in the constraint.

\noindent\textit{\textbf{Conclusion---}} In this work, we have modeled the effects of NVNL for a BBH merger and estimated the detectability in current and third generation detectors. By incorporating NVNL fluctuations to an EOB model, we obtained modified trajectories for the binary, and the corresponding gravitational waveforms. These waveforms' phase deviation from GR in the frequency domain can be well approximated as a single mode function times a random coefficient with normal distribution with zero mean and standard deviation $A$, which in turn also characterizes the typical size of the metric perturbations close to the horizon. We estimated constraints that can be posed by LVK and third-generation detectors. We showed that the ``optimal'' PCA templates constrains $A$ tighter than the standard PN parameters by about 30\%. To perform our analysis on actual data, we need to take the further step of making an NVNL-EOB model which includes spin effects and is properly calibrated to numerical relativity waveforms, for example with SEOBNRv5 \cite{Pompili:2023tna}.

This work is primarily concerned with the finding a qualitative picture of waveforms in NVNL. Since we are particularly concerned with the behavior of comparable mass systems like LVK detects, we tuned our approach accordingly and used the EOB model. Thus, our work has been limited to conservative dynamics during the inspiral stage. To obtain a more complete picture of NVNL effects, one should: (i) combine our work with Ref.~\cite{Fransen:2024fzs} to incorporate NVNL's modifications to radiation reaction during the inspiral stage, and (ii) model the effect of NVNL perturbations on wave propagation through the final BH's spacetime in order to capture how the ringdown waves will be modified. We stress that modeling the dissipative and ringdown effects is particularly difficult in the comparable mass case since the EOB modeling choice is not a fully first-principles approach. While there is no particular reason that the conservative effects are dominant, these results are a first order-of-magnitude picture for NVNL in ground based detectors.

\noindent \textit{\textbf{Acknowledgments---}} We thank Jacob Golomb, Ethan Payne, Sophie Hourihane, Max Isi, Bangalore Sathyaprakash, Matthew Giesler, Steve Giddings, Hang Yu, and Katerina Chatziioannou for their insightful discussions. B.S. acknowledges support by the National Science Foundation Graduate Research Fellowship under Grant No. DGE-1745301. Y.C. and B.S. acknowledge support from the Brinson Foundation, the Simons Foundation (Award No.~568762), and by NSF Grants No.~PHY-2309211 and No.~PHY-2309231.

\appendix
\begin{center}
    \textbf{Supplemental Material}
\end{center}

\noindent \textit{\textbf{EOB Waveform---}} We made our EOB waveform by modifying the nonspinning EOBNRv2 waveform \cite{Pan:2011gk}. If we parameterize our metric with coordinates $\vec q = (r,\phi)$ and their conjugate momenta $\vec p = (p_r,p_\phi)$, the effective Hamiltonian is found with the mass shell condition $p_\mu p_\nu g^{\mu\nu}_\mathrm{S} = - 1$ 
\begin{equation}
    \hat H_\mathrm{eff}^\mathrm{S} = \sqrt{A(r) \left( 1 + \frac{p_\phi^2}{r^2} + \frac{A}{D} p_r^2 \right)} \, .
\end{equation}
The physical Hamiltonian for the system is related to the effective Hamiltonian via 
\begin{equation}
    \hat H_\mathrm{real}^\mathrm{S} = {\eta}^{-1}\sqrt{1 + 2 \eta \left(\hat H_\mathrm{eff} - 1 \right)} \, ,
\end{equation}
where the hat denotes the Hamiltonian is in dimensionless units. Hamilton's equations are then 
\begin{equation}\label{eq:hamiltons}
        \frac{d q_i}{dt} = \frac{\partial \hat H_\mathrm{real}}{\partial p_i} \, ,\quad 
        \frac{d p_i}{dt} = -\frac{\partial \hat H_\mathrm{real}}{\partial q_i}  +  \mathcal{F}^{\rm rad}_i \, ,
\end{equation}
where the generalized force $\mathcal{F}_i^{\rm rad}$ is added to incorporate radiation reaction. The reason that we can add NVNL directly to the EOB spacetime is that this is the leading order term, and all other terms will be smaller by a factor of the symmetric mass ratio $\eta \leq 1/4$ (so our calculations are accurate to $\mathcal{O}( A\cdot \eta)$). Thus for more asymmetric systems this approximation improves.

Given an effective metric, one can solve for the effective Hamiltonian by solving the mass-shell constraint $p_\mu p_\nu g^{\mu\nu} = -1 - Q(p^4)$. If perturbations from NVNL are added this is generically
\begin{widetext}
    \begin{equation}\label{eq:perturb-heff}
        \hat H^\mathrm{eff}(r,\phi, p_r, p_\phi) =  \frac{g^{0i} p_i}{g^{00}}  + \sqrt{\frac{1 + g^{ij} p^i p^j }{- g^{00}} + \left( \frac{g^{0ri} p_i}{g^{00}} \right)^2 + \frac{Q(p^4)}{-g^{00}} } \, ,
    \end{equation}
\end{widetext}
where the modification to the mass shell is $Q(p^4) = 2\eta \left( 4-3\eta \right) p_r^4$ which we do not modify due to NVNL since it is a high PN order effect. By perturbing $g^{\mu\nu}$ in Eq.~\eqref{eq:perturb-heff}, we find the perturbation to the real Hamiltonian $\Delta \hat H^{\ell m}$ 
\begin{equation}
    \hat H_\mathrm{real} = \hat H_\mathrm{real}^\mathrm{S} + h_{vv}^{\ell m} \, \Delta \hat H_{\ell m}^\mathrm{real} \, .
\end{equation}
In this section, we will be using dimensionless units so $t = T/M$, $r = R/M$, $p_r = P_R/M$, and $p_\phi = P_\phi / (M\mu)$.
The equations of motion for the EOB trajectory are found by solving Hamilton's equations with radiation reaction terms. If we explicitly write out the EOB trajectory evolution equations with the perturbations we have 
\begin{align}
    \frac{\partial  r}{\partial t} &= \frac{\partial \hat{H}^\mathrm{S}_\mathrm{real}}{\partial p_r} 
    +  \frac{\partial \Delta \hat H_{\ell m}^\mathrm{real}}{\partial p_r} h_{vv}^{\ell m} \, , \nn\\
    \frac{\partial \phi}{\partial  t} = \hat \omega &= \frac{\partial \hat H_\mathrm{real}^\mathrm{S}}{\partial  p_\phi} 
    + \frac{\partial \Delta \hat H_{\ell m}^\mathrm{real}}{\partial  p_\phi} h_{vv}^{\ell m} \, ,\nn \\
    \frac{\partial  p_r}{\partial  t} &= -\frac{\partial \hat H_\mathrm{real}^\mathrm{S}}{\partial r} + \hat{\mathcal{F}}_\phi \frac{ p_r}{ p_\phi} 
    -\frac{\partial \left(\Delta \hat H_{\ell m}^\mathrm{real} h_{vv}^{\ell m}\right)}{\partial r}  \, , \nn\\
    \frac{\partial p_\phi}{\partial t} &= \hat{\mathcal{F}}_\phi 
    -\frac{\partial \left(  h_{vv}^{\ell m}\right)}{\partial \phi} \Delta \hat H_{\ell m}^\mathrm{real}\, , \label{eq:perturbation-eob}
\end{align}
and the $\phi$ component of the radiation-reaction force is 
\begin{equation}
    \hat{\mathcal{F}}_{\phi}=-\frac{1}{\eta v_{\omega}^3} \frac{d E}{d t} \, ,
\end{equation}
where $v_{\omega} = \hat \omega^{1/3}$. The GW luminosity is generally 
\begin{align}
    \frac{d E}{d t}=\frac{v_{\omega}^6}{8 \pi} \sum_{\ell=2}^\infty \sum_{m=\ell-2}^{\ell} m^2\left|\frac{D_L}{M} h_{\ell m}\right|^2 \,,
\end{align}
however we make the approximation and only include the $(2,2)$ mode. To construct $h_{22}(t)$, we are using the Newtonian contribution as given in Eq.~(16) of \cite{Pan:2011gk}. It is equal to 
\begin{equation}
    h_{22} =  - \frac{32 \pi}{5} \sqrt{\frac{2}{3}} \frac{M \eta}{D_L} v_\phi^2 Y^{2-2}\left( \frac{\pi}{2}, \phi \right) \, ,
\end{equation}
where
\begin{equation}
    v_{\phi} \equiv \hat{\omega} r_{\omega} \equiv \hat{\omega} r\left[\psi\left(r, p_{\phi}\right)\right]^{1 / 3} \, ,
\end{equation}
and 
\begin{equation}
    \psi\left(r, p_{\phi}\right)=\frac{2\left\{1+2 \eta\left[\sqrt{A(r)\left(1+p_{\phi}^2 / r^2\right)}-1\right]\right\}}{r^2 d A(r) / d r} .
\end{equation}
Since our analysis is focusing on how the waveform differs by adding metric perturbations away from GR, we neglected to include various calibration terms that are included in EOBNRv2. In Ref.~\cite{Pan:2011gk}, they use the factorized resummed modes \cite{Damour:2008gu,Damour:2009kr,Buonanno:2009qa} which include corrections to the Newtonian modes motivated from numerical relativity. Specifically, we use $h_{\ell m}^F = h_{\ell m}^N$ in Eq.~(14) of \cite{Pan:2011gk}. We also do not include the effects of the non-quasicircular orbit coefficients in Eq.~(13) of \cite{Pan:2011gk}.

We attach a phenomenological ringdown to our waveform. This is necessary so that the waveform can have a well defined Fourier transform / stationary phase approximation $h(f)$ which is well defined when a NVNL merges before the GR one. We stress that this is only used when extracting $\Delta \Psi(f)$, and the parameter estimation in the main work uses IMRPhenomD \cite{Husa:2015iqa,Khan:2015jqa} with extra beyond GR phase $\Delta \Psi(f)$. Since we did not include the GR calibration from the non-quasicircular orbit coefficients, we found that it was hard to get a good fit with the comb approach used in \cite{Pan:2011gk}. This is because the quasinormal modes are at a much higher frequency than the gravitational wave frequency at the merger-ringdown fit point. Without properly fitting to numerical relativity, we saw a preference for unphysical second peaks similar to what is shown in Fig.~3 of \cite{Pan:2011gk}. Instead, we choose the fit location $\partial_t |h_{22}|=0$ and attach a ringdown. If we write the $h_{22}$ waveform as 
\begin{equation}
    h_{22}(t) = A_{22}(t) e^{- i \phi_{22}(t)} \, ,
\end{equation}
where time is scaled so that merger happens $t=0$. The phenomenological ringdown waveform is described by
\begin{equation}
    \omega_{22,\mathrm{RD}}^r(t) = \omega_{220}^r \left(1-\sum_{i=1}^2 \alpha_i e^{-t/\tau_i} \right)\, ,
\end{equation}
and the ringdown amplitude is 
\begin{equation}
    A_{22,\mathrm{RD}}(t) = A_{22}|_{t=0}  \exp\left[\omega_{220}^i t \left(1 + \sum_{i=1}^2\beta_i e^{-t/\tau_i} \right)\right] \, ,
\end{equation}
where $\omega_{220} = \omega_{220}^r + i\, \omega_{220}^i$. We set $\alpha_i$ and $\beta_i$ by enforcing that the match between inspiral and ringdown is twice continuously differentiable $h_{22,\mathrm{IM}}'(t) = h_{22,\mathrm{RD}}'(t)$ and $h_{22,\mathrm{IM}}''(t) = h_{22,\mathrm{RD}}''(t)$. We note that the accuracy of the ringdown fit isn not that important because we are only searching for deviations from GR during the inspiral-merger and assume $\Delta \Psi(f)$ is constant after merger (as can be seen in Fig.~\ref{fig:phase-shift} above $M f\sim 0.4$). We stress that attaching a GR ringdown is a conservative choice, and suspect that proper modeling of the ringdown could improve constraints by a factor of $\sim2$. Our analysis of NVNL in the ringdown phase is part of an ongoing future work.

\noindent\textit{\textbf{Analytical Description---}} In this section, we will describe how the coordinates are modified in the inspiral and explain the perturbative framework in more detail. Let us denote the EOB state space coordinate as $x^a(t) = \left( r, \phi,p_r,p_\phi \right)$ and consider deviations away from the trajectory that the GR waveform takes $x^a_\mathrm{GR}(t)$. The deviations $\Delta x^a(t) = x^a_\mathrm{NVNL}(t)-x^a_\mathrm{GR}(t)$ follow a coupled system of differential equations
\begin{equation}
    \frac{d \Delta x^a}{dt} = M^{ab}(t)\Delta x_b + F^a(t) \,,
\end{equation} 
where $M^{ab}(t)$ represents perturbations away from the GR trajectory associated with perturbations of GR terms $\hat H_\mathrm{real}^\mathrm{S}/\hat{\mathcal{F}}_\phi$ in Eq.~\eqref{eq:perturbation-eob} while $F^a(t)$ is the original sourced deviation from terms containing $h_{vv}^{\ell m}$ in Eq.~\eqref{eq:perturbation-eob}. Typically, $M^{ab}(t)$ is either constant--e.g.\ in a simple harmonic oscillator--or features damping when friction is present. In contrast, $M^{ab}(t)$ provides an anti-restoring feedback that causes perturbations to grow secularly. Consequently, although NVNL terms only initially source small deviations in the EOB equations of motion, the orbital dynamics secularly amplify these deviations over time. Physically, this can be understood as a slight eccentricity induced by the NVNL forces, which shortens/lengthens the inspiral and leads to an earlier coalescence (depending on sign of angular momentum kick).

Additionally, it is important to note how the frequency domain dephasing is related to the changes from NVNL waveform. If our waveforms are of the form $h_\mathrm{GR}(t) = A_\mathrm{GR}(t) e^{-2\phi_\mathrm{GR}(t)}$, the frequency domain dephasing in the stationary phase approximation is equal to
\begin{align}
    \Delta \Psi(f) &= 2 \pi f \Delta t(f) - 2 \left[ \phi_\mathrm{nvnl}(t_\mathrm{nvnl}(f)) - \phi_\mathrm{GR}(t_\mathrm{GR}(f)) \right]\,, \nonumber \\
    &= - 2 \Delta \phi(f)\,,
\end{align}
where $\Delta \phi(f)=\phi_\mathrm{nvnl}(t_\mathrm{GR}(f)) - \phi_\mathrm{GR}(t_\mathrm{GR}(f))$ is the time domain orbital phase deviation. Therefore, deviations to orbital phase in the inspiral are directly related to the frequency domain phasing in the stationary phase approximation when appropriately using the time frequency relation (also true in ringdown).

\begin{figure}[ht]
    \centering
    \includegraphics{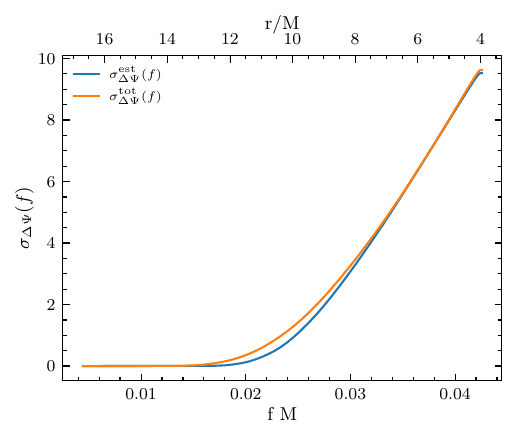}
    \caption{Comparison of how much variance is captured by the PCA estimator. We see that the full $\Delta\Psi$ is well described by this. While some of the variance is not captured at low frequencies, at high frequencies there is a nearly perfect match, especially at larger frequencies. The other PCA terms contain about $3\%$ of the variance that is not accounted for in the primary one here as discussed in the main text.}
    \label{fig:psi-alpha-correlation-extraplot}
\end{figure}

\noindent\textit{\textbf{Accuracy of PCA---}}
Let us now discuss the accuracy of the PCA. We do this by comparing the variance in the phase deviation that is captured by our PCA model to the full phase deviation. We find the full dephasing by computing a FFT $\Delta \Psi_\mathrm{full}(f)$ and the PCA dephasing is $\Delta \Psi_\mathrm{PCA}(f) = \zeta z(f)$ where $z(f)$ is the most dominant principal mode and $\zeta$ is a parameter so that $|\Delta \Psi_\mathrm{full}(f) - \Psi_\mathrm{PCA}(f)|$ is minimized. The variance captured at each frequency $f$ by the PCA is
\begin{align}
    \sigma_{\Delta\Psi}^\mathrm{est}(f) &\equiv \frac{\langle \Delta\Psi_\mathrm{PCA}(f) \, \Delta \Psi_\mathrm{full}(f)\rangle}{\sqrt{\langle \Delta\Psi_\mathrm{PCA}^2(f)  \rangle }} \, ,\nn \\
    &=\frac{\langle \zeta \, \Delta \Psi_\mathrm{full}(f)\rangle}{\sqrt{\langle \zeta^2  \rangle }} \, .
\end{align}
This needs to be compared to total amount of variance in the full waveform
\begin{equation}
    \sigma_{\Delta\Psi}^\mathrm{tot}(f) = \sqrt{\langle \Delta \Psi_\mathrm{full}^2(f) \rangle} \, .
\end{equation}
In Fig.~\ref{fig:psi-alpha-correlation-extraplot}, we compare these variance indicators and see the fit quality. One can see that the variance captured by the PCA estimator $\Psi_\mathrm{PCA}(f)$ is less than the true variance in $\Delta \Psi_\mathrm{full}(f)$, however it does a good job of estimating the variance at high frequencies when the dephasing is largest. Note that the reason that these do not perfectly match up is that the PCA is optimizing the quantity $|\Sigma(f,f') - \sigma^2 z(f) z(f')|$. The off diagonal elements of $\Sigma(f,f')$ where $f\ne f'$ are better fit by choosing $z(f)$.

\begin{figure}[ht]
    \centering
    \includegraphics{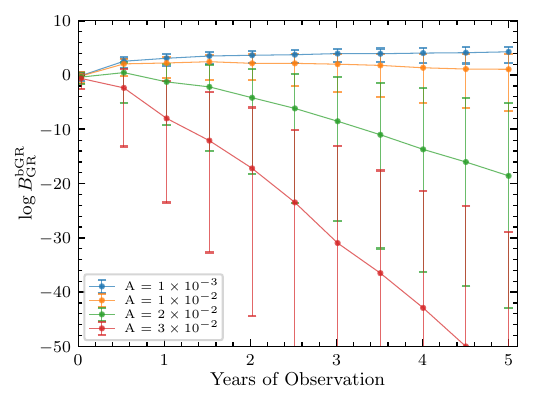}
    \caption{The log Bayes factor for the Hanford-Livingston-Virgo network at O3 Livingston sensitivity (positive favors GR). One can see that the red curve favors NVNL at large events, while blue favors GR. The whiskers correspond to the upper and lower bounds are $\pm 1\sigma$ percentile values due to randomness associated with the order of events. If one sees a loud clear event early, then it is easier to favor/disfavor GR.
    }
    \label{fig:savage-dickey-O3-range}
\end{figure}

\noindent \textit{\textbf{Variance of Bayes Factor---}}
As we noted in the discussion of Fig.~\ref{fig:bayes-factor-result-O3} and Fig.~\ref{fig:bayes-factor-result-CE}, the Bayes factor measurement depends on randomness about the event order. In Fig.~\ref{fig:savage-dickey-O3-range}, we plot how the (log) Bayes Factor ratio scales for multiple injection sizes of $A$. In the center, we show the median Bayes factor for each injection size while the upper and lower bounds are $\pm 1\sigma$ percentile values for the Bayes factor after this many observations. This mostly occurs because the loudest events are the most informative, so the order of events can affect the rolling constraint. Furthermore, the statistical realization of the detector noise and hierarchical model draw add subdominant variations to the Bayes factor.

\begin{figure}[ht]
    \centering
    \includegraphics{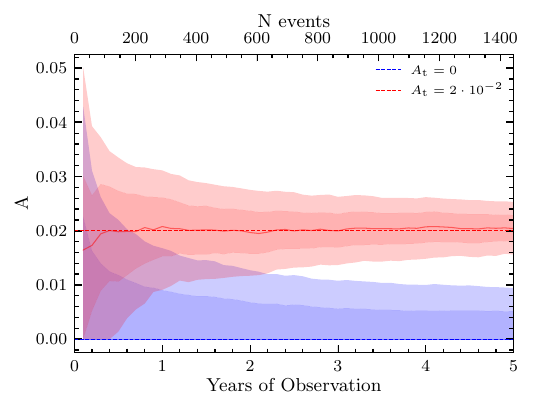}
    \caption{We show the running estimate of $A$ for an O3 network. In blue, we show the $1\sigma$ and $2\sigma$ confidence region when we inject $A_\true=0$. We can see that the error shrinks as events are observed. In blue, we inject $A_\true = 2\times 10^{-2}$, and one can see that as events are observed the probability density narrows and detects a violation.}
    \label{fig:1d-LIGO-consistency}
\end{figure}
\begin{figure}[ht]
    \centering
    \includegraphics{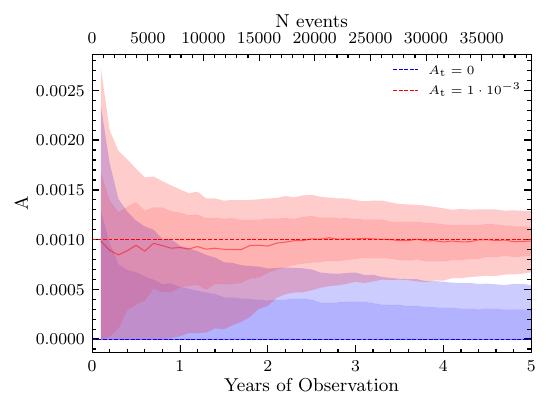}
    \caption{We show the running estimate of $A$ for the CE network in an analogous manner of Fig.~\ref{fig:1d-LIGO-consistency}. In red, we inject $A_\true=0$, and in blue $A_\true = 10^{-3}$. One can see that as more observations are made, the confidence region begins to exclude $A = 0$.}
    \label{fig:1d-CE-consistency}
\end{figure}

\noindent \textit{\textbf{Measuring $A$---}} While in the main text of this letter, we primarily focused on null tests of NVNL, one can directly measure the size of the deviations. In the same manner that we did before, we can compute the event posteriors for many events and then compute the posterior in the hierarchical model. The true value of NVNL parameter $A$ is for the choice of $p(\mu,\sigma|d) = p(0,A|d)$. We can thus compute the confidence interval on $A$ by finding the maximum posterior location and shade the interval which contains 1$\sigma$ and 2$\sigma$ of the posterior support. We do this for a LIGO-Virgo network at O3 Livingston sensitivity in Fig.~\ref{fig:1d-LIGO-consistency}, where in blue is an injection of $A_\true = 0$, and $A_\true = 2\times 10^{-2}$ constraints are shown in red. We shade the $\pm 1\sigma$ and $\pm 2\sigma$ regions of the posterior as a function of the years of observation. One can see that the case of zero injection slowly asymptotes to a stronger constraint on $A$ while injection a nonzero $A$ the credible interval eventually detects it at 95\% level after a little less than a year. In Fig.~\ref{fig:1d-CE-consistency}, we plot the same case for CE with $A_\true = 0$ in blue and $A_\true = 10^{-3}$ in red. One can see similar features where the constraint narrows down to the true value as the network observes more events.

\FloatBarrier 
\bibliography{bibliography.bib}

\end{document}